\documentclass[preprint,3p,times,twocolumn]{elsarticle} 
\usepackage{stmaryrd} 

 \usepackage{amssymb}
 \usepackage{graphicx}
 \usepackage{amsmath} 
 \usepackage{color}
 \usepackage{subfigure}
 \usepackage{hyperref}
 \usepackage{pdflscape}
 \usepackage{bm}
 \usepackage[page]{appendix}
 \usepackage{multirow}

 \newcommand{\Fig}[1]{Figure~\ref{fig:#1}}
 
 \newcommand{\Eqn}[1]{Equation~(\ref{eq:#1})}
 \newcommand{\Eqnlist}[2]{Equations~(\ref{eq:#1}) and (\ref{eq:#2})}
 \newcommand{\Eqnset}[1]{Equations~(\ref{eq:#1})}
 \newcommand{\Eqnrange}[2]{Equations~(\ref{eq:#1}\,-\,\ref{eq:#2})}
 \newcommand{\Sec}[1]{Section~\ref{sec:#1}}
 
 \newcommand{\Tab}[1]{Table~\ref{tab:#1}}
 \newcommand{\etal}{\emph{et~al}\ }

 \renewcommand{\vec}[1]{\mbox{$\bf #1$}} 
 
 \newcommand{\dotv}{  \mbox{\,\boldmath\(\cdot\)\,} }
 \newcommand{\iotabar}{\ensuremath{\,\iota\!\!}\textrm{-}}     

 \newcommand{\const}{\mbox{const}}
 \newcommand{\curl}{\nabla \times}
 
 \newcommand{\cross}{\times}
 \renewcommand{\div}{\nabla \cdot}
 
 \newcommand{\deriv}[2][\null]{\frac{d#1}{d#2}}

 \newcommand{\pderiv}[2][\null]{\frac{\partial #1}{\partial #2}}

 \newcommand{\jump}[1]{\left[ \!\!\left[ #1 \right] \!\! \right]}
 \newcommand{\smljump}[1]{\left[ \!\left[ #1 \right] \! \right]}

\newcommand{\grad}{\bm{\nabla} }
\newcommand{\divv}{\bm{\nabla\cdot} }
\renewcommand{\dotv}{\bm{\cdot} }
\renewcommand{\cross}{\bm{\times} }
\renewcommand{\curl}{\bm{\nabla\times} }

\renewcommand{\jump}[1]{\left\llbracket  #1 \right\rrbracket} 
\newcommand{\Id}{\textsf{\textbf{I}}} 
\newcommand{\T}{\textsf{\textbf{T}}} 

\renewcommand{\Fig}[1]{Fig.~\ref{fig:#1}}
\renewcommand{\Eqn}[1]{Eq.~(\ref{eq:#1})}

\renewcommand{\Sec}[1]{Sec.~\ref{sec:#1}}

\renewcommand{\etal}{\emph{et~al.}}

\begin{document}

\begin{frontmatter}
\title{Hamilton--Jacobi theory for continuation of magnetic field across a toroidal surface supporting a plasma pressure discontinuity}
\author{M. McGann}
\ead{mathew.mcgann@anu.edu.au}
\address{Plasma Research Laboratory, Research School of Physics \& Engineering, The Australian National University, Canberra, ACT 0200, Australia}
\author{S.R. Hudson}
\address{Princeton Plasma Physics Laboratory, PO Box 451, Princeton NJ 08543}
\author{R.L. Dewar}
\address{Plasma Research Laboratory, Research School of Physics \& Engineering, The Australian National University, Canberra, ACT 0200, Australia}
\author{G. von Nessi}
\address{Plasma Research Laboratory, Research School of Physics \& Engineering, The Australian National University, Canberra, ACT 0200, Australia}
\date{\today}

\begin{abstract}

The vanishing of the divergence of the total stress tensor (magnetic plus kinetic) in a neighborhood of an equilibrium plasma containing a toroidal surface of discontinuity gives boundary and jump conditions that strongly constrain allowable continuations of the magnetic field across the surface. The boundary conditions allow the magnetic fields on either side of the discontinuity surface to be described by surface magnetic potentials, reducing the continuation problem to that of solving a Hamilton--Jacobi equation. The characteristics of this equation obey Hamiltonian equations of motion, and a necessary condition for the existence of a continued field across a general toroidal surface is that there exist invariant tori in the phase space of this Hamiltonian system. It is argued from the Birkhoff theorem that existence of such an invariant torus is also, in general, sufficient for continuation to be possible.  An important corollary is that the rotational transform of the continued field on a surface of discontinuity must, generically, be irrational.

\end{abstract}


\begin{keyword}
Hamiltonian dynamics \sep invariant tori \sep pressure discontinuities \sep plasma 
\end{keyword}

\end{frontmatter}

\maketitle

\section{Introduction}\label{sec:Intro}

This paper is concerned with plasma confinement in nonaxisymmetric toroidal magnetic fields, such as occur in fusion experiments. As is typical in plasma equilibrium theory we assume the plasma to be static (i.e. mass flow is negligible) and the ion and electron Larmor radii to be negligible. There is no minimum scale length in this model and discontinuities are in principle possible, provided the net force density at each point $\vec{r}$ in the plasma vanishes (the \emph{force balance condition}). Also, a corollary of the assumed flowless state is that electric fields are also negligible. 

We assume the kinetic stress is described by the isotropic pressure tensor $P(\vec{r})\Id$, where $\Id$ is the unit dyadic. Adding the electromagnetic stress, the total stress tensor is
\begin{align}\label{eq:stresstensor} 
	\T \equiv P\, \Id + \frac{B^2}{2} \Id - \vec{B}\vec{B}\;,
\end{align}
where $\vec{B} \equiv \vec{B}_{\rm SI}(\vec{r})/\sqrt{\mu_0}$, $\vec{B}_{\rm SI}$ being the magnetic field in SI units and $\mu_0$ the permeability of free space. The force balance condition is then~\cite{Freidberg1987}
\begin{align}\label{eq:stressbalance} 
	\divv\T = 0\;.
\end{align}

If $P$ and $\vec{B}$ are differentiable, \Eqn{stressbalance} may be written in the more usual form $\grad P = (\curl\vec{B})\cross\vec{B}$; but at surfaces of discontinuity $P$  and $\vec{B}$ are step functions and $\curl\vec{B}$ contains a Dirac delta function component, corresponding to a sheet current at the discontinuity. Then $(\curl\vec{B})\cross\vec{B}$ is not well defined and it is better to regard \Eqn{stressbalance}, interpreted in the weak sense of distribution theory, as the fundamental form of the force balance relation.

Magnetic field lines can be regarded as trajectories of a $1\frac{1}{2}$ degree-of-freedom Hamiltonian dynamical system~\cite[e.g.]{Boozer2004}, with the toroidal angle $\zeta$ taken to be the analog of time and the half degree of freedom implying the system is not, in general, autonomous---the Hamiltonian depends explicitly on $\zeta$. An important exceptional case is that of axisymmetry, as in an ideal tokamak, in which case the field-line Hamiltonian is integrable and all field lines lie on invariant tori foliating the plasma volume. It is then consistent with \Eqn{stressbalance} to assume $P$ and $\vec{B}$ to be arbitrarily smooth functions, with the isosurfaces of $P$ coinciding with the invariant tori.

In the theory of magnetically confinement plasmas these invariant tori are called \emph{magnetic surfaces}, and each is characterized by a field-line winding number called the \emph{rotational transform} and denoted $\iotabar$, being the asymptotic limit of the number of poloidal transits divided by the number of toroidal transits as the length of a field-line segment is taken to infinity.

However, in stellarators and real tokamaks axisymmetry is (to a greater or lesser extent) broken and the magnetic field is nonintegrable. Then, while magnetic surfaces of sufficiently irrational $\iotabar$, may exist [by the Kolmogorov--Arnold--Moser (KAM) theorem for small departures from axisymmetry, or by good design in systems far from axisymmetry], the intervening magnetic surfaces are broken and replaced by a fractal structure of island chains, which include periodic orbits and chaotic regions.

In this case the problem of finding $P$ and $\vec{B}$ globally consistent with \Eqn{stressbalance} is notoriously difficult~\cite{Grad67}, and it is an open question as to whether solutions with globally continuous but nonconstant $P$ exist. (In the linear force-free case of \emph{constant} $P$ throughout the plasma, when $\vec{B}$ obeys the Beltrami equation, $\curl\vec{B} = \mu\vec{B}$, with $\mu = \const$, solutions are known to exist under appropriate boundary conditions \cite{Kress1981,Yoshida1990}, irrespective of the integrability of $\vec{B}$.) However, by appeal to KAM theory it has been shown~\cite{B&L} that discontinuous solutions with piecewise constant $P$ exist for small departures from axisymmetry, and recent work~\cite{HHD07,Dewar2008} holds out promise that variational methods can be used to construct numerical solutions with a stepped pressure profile for systems far from axisymmetry.

In the present paper we consider a simpler, more local \textbf{problem}: Consider a \emph{prescribed} torus $\cal S$ and assume $P$ and $\vec{B}$ differentiable in open regions bounded by $\cal S$, with $P$ constant on each side of $\cal S$ but discontinuous across it. Then, given the field $\vec{B}(\vec{r})$ on one side of $\cal S$ and $\iotabar$ on the other side, find the field $\vec{B}(\vec{r})$ on the other side that is consistent with force balance in the weak sense discussed above.

In \Sec{Formulation} we show that the surface magnetic fields must be derivable from potential functions defined on either side of the surface, so that the prescribed $\vec{B}$ can be obtained by specifying a scalar function.  The unknown surface potential on the other side can be determined by solving a Hamilton--Jacobi equation involving a Hamiltonian, the \emph{pressure jump Hamiltonian}. The configuration space on which this Hamiltonian is defined corresponds to a \emph{single} flux surface, while the ``momenta'' correspond to possible magnetic fields on the same surface. Thus the pressure jump Hamiltonian is quite distinct from the magnetic field line Hamiltonian described above, which is defined globally. (These Hamiltonians are compared and contrasted in Appendix B. The only relevance of the magnetic field line Hamiltonian to the problem posed in this paper is that $\cal S$ must be an invariant torus of the field line Hamiltonian dynamics.)

Earlier, Berk \etal~\cite{Berk1986} introduced a Hamilton--Jacobi approach to investigate an individual flux surface's ability to withstand pressure, but their scope was limited to the case of vanishing current field on one side of the surface and vanishing magnetic field on the other.  Kaiser and Salat~\cite{K&S} attempted to prove the existence of flux surfaces using only geometrical arguments (geodesic flow), but this approach assumed the same field simplifications as Ref.~\cite{Berk1986}. By contrast our formulation allows nonzero current and magnetic field on both sides of $\cal S$.

The geodesic approach of Ref.~\cite{K&S} was introduced because of a worry that points on an invariant torus embedded in the phase space of the pressure jump Hamiltonian may not map one-to-one to the configuration space of the torus, violating the condition that field lines cannot cross.  Kaiser and Salat were also doubtful of the applicability of the KAM theory (which had been alluded to in Ref.~\cite{Berk1986}) to ensure the existence of some flux surfaces under small deformations. 

In \Sec{Analytical} we demonstrate the applicability of the Hamiltonian formulation by applying the Birkhoff theorem to prove the existence of a mapping that takes characteristics on invariant tori in the phase space of the pressure jump Hamiltonian to the magnetic field lines in configuration space, such that the topology of those characteristics is preserved. 
We then briefly discuss conditions and methods for finding such invariant tori, arguing that, while KAM theory is applicable to systems close to axisymmetry, this is too restrictive to be of much practical use, whereas Greene's residue method is practical if not completely rigorous.

One might be tempted to infer from this, and the fact that invariant tori in nonintegrable Hamiltonian systems generically have irrational winding numbers, that $\iota$ must be irrational on both sides of $\cal S$. However, in the plasma equilibrium problem, the field line dynamics is not known \emph{a priori} but rather to be found by solving for $\vec{B}$ self-consistently, with a boundary condition being that $\vec{B}$ be tangential on the two sides of $\cal S$. Thus, it could be said that $\cal S$ is an invariant torus of the field line dynamics \emph{by construction}. It is one of the main aims of the present paper to show that force balance across $\cal S$ provides an independent, locally self-consistent reason for believing that $\iota$ must be irrational on both sides of $\cal S$ in almost all nonaxisymmetric equilibria.

\section{Formulation}\label{sec:Formulation} 

\subsection{Physical Derivation}


We are interested in weak solutions of the coupled pair of equations, \Eqn{stressbalance} and the divergence-free magnetic field condition,
\begin{align}\label{eq:div} 
 \div \vec{B} = 0\;,
\end{align}
in the neighborhood of a given surface of discontinuity.

Much can be inferred from \Eqn{stressbalance}; the details are in
Appendix~\ref{app:presjump}
but here we report the important results.  When there is a discontinuity in the pressure across a surface $\mathcal{S}$, and when the magnetic field is force free in the neighborhood of the surface, the following must apply on $\cal S$,
\begin{subequations}\label{eq:existence}
 \begin{align}
                                P &= \textrm{const} \;,\label{eq:Pconst} \\
            \vec{B} \dotv \vec{n} &= 0\;,\label{eq:Btangent} \\
  ( \curl \vec{B} ) \dotv \vec{n} &= 0\;,\label{eq:Bcrosstangent}  \\
       \jump{\frac{1}{2} B^2 + P} &= 0\;.\label{eq:forcebalance} 
 \end{align}
\end{subequations}
In \Eqnset{existence}, $\vec{n}$ is the normal to the surface, and the notation $\smljump{x}$ refers to the difference between $x$ on one side of the surface and $x$ on the other side.  \Eqnrange{Pconst}{Bcrosstangent} must hold on both sides of the surface [with a different constant on either side in the case of \Eqn{Pconst}].  We refer to \Eqnset{existence} collectively as the \emph{pressure jump conditions}.  Note the tangential component of the magnetic field is discontinuous.  

To describe toroidal fusion geometries we now consider the surface $\mathcal{S}$ to have a toroidal topology and use the coordinate system $(\theta,\zeta)$.  Here, $\theta$ and $\zeta$ are angle-like coordinates in the poloidal and toroidal directions respectively.  

\begin{figure}
\centering
\includegraphics[width=0.4\textwidth]
 {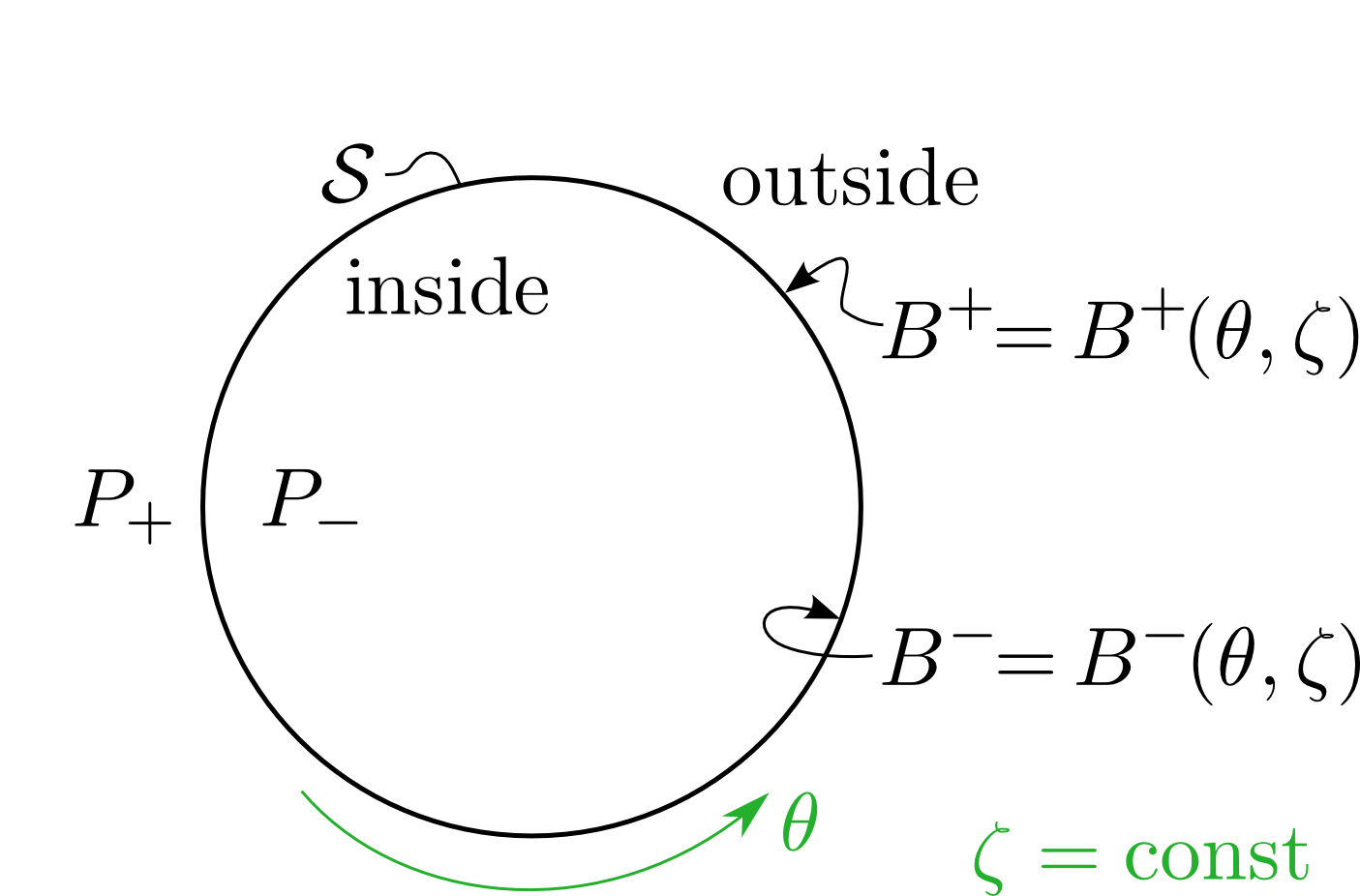}
\caption{(color online) A toroidal cut showing the cross section of the plasma and the various definitions.  The cross section is a circle for simplicity, but the theory will work for a surface $\mathcal{S}$ of any shape.  }
\label{fig:2Dsetup} 
\end{figure} 

The field on the inner side (side closest to the magnetic axis) of $\mathcal{S}$ is given by $\vec{B}^-$, and the field on the outer side is given by $\vec{B}^+$, as shown in \Fig{2Dsetup}.  Equation~(\ref{eq:forcebalance}) then becomes
\begin{align}\label{eq:simpjump}
 {B^+}^2 - {B^-}^2 = 2(P_- - P_+)\;.
\end{align}

When \Eqn{simpjump} is written in terms of the covariant components of the magnetic fields, one finds that
\begin{align}\label{eq:our7}
 2 \Delta P = \sum_{i,j \in \{\theta,\zeta\}} g^{ij} \left[B^+_i B^+_j - B^-_i B^-_j \right]\;,
\end{align}
where $\Delta P = P_- - P_+$ and $g^{ij} = g^{ij}(\theta,\zeta)$ are the contravariant metric coefficients defined on the 2-dimensional Riemannian manifold $\mathcal{S}$, related to the covariant coefficients~\cite{B&L} by inverting the $2\times 2$ matrix $[g_{ij}]$,
\begin{align}
g^{\theta\theta} &=   g_{\zeta  \zeta  }/\sqrt{g}\;, &
g^{\theta\zeta  } &= - g_{\theta\zeta  }/\sqrt{g}\;, &
g^{\zeta  \zeta  } &=   g_{\theta\theta}/\sqrt{g}\;,
\end{align} 
where $\sqrt{g} = g_{\theta\theta} g_{\zeta\zeta} - g_{\theta\zeta}^2$ is the determinant.  We shall also need the 2-dimensional contravariant components of the surface magnetic field, defined by
\begin{subequations}\label{eq:Bcocontra}
\begin{align} 
 B^\theta &= g^{\theta\theta} B_\theta + g^{\theta\zeta} B_\zeta\\
 B^\zeta   &= g^{\theta\zeta} B_\theta + g^{\zeta\zeta} B_\zeta\;.
\end{align}
\end{subequations}

The torus $\mathcal{S}$ is embedded in an ambient, 3-dimensional Euclidean space with its own metric, $G^{ij}$ say.  If $(\theta,\zeta,\psi)$ are curvilinear coordinates in this ambient space such that $\psi = \const$ on $\mathcal{S}$, then the covariant components $G_{ij}$, $i,j\in\{\theta,\zeta\}$, are identical to $g_{ij}$ on $\mathcal{S}$, but the contravariant components are different,
\begin{align}
 g^{ij} = G^{ij} - \frac{G^{\psi i}G^{\psi j}}{G^{\psi\psi}} \:\mbox{for}\: i,j\in\{\theta,\zeta\} \;.
\end{align}

\subsection{Hamiltonian Treatment}\label{sec:PJH} 

The keys to the Hamiltonian treatment are \Eqnlist{Btangent}{Bcrosstangent}, together they imply that on $\mathcal{S}$, 
\begin{align} \label{eq:allowf} 
 \partial_\theta B^{\pm}_\zeta - \partial_\zeta B^{\pm}_\theta = 0\;.
\end{align}
\Eqn{allowf} is implicitly satisfied if the field components are written as
\begin{align}\label{eq:fdef} 
 B^{\pm}_\theta &= \partial_\theta f^{\pm}\;, & B^{\pm}_\zeta &= \partial_\zeta f^{\pm}\;,
\end{align}
where the two scalar functions $f^{\pm}(\theta,\zeta) = \int \vec{B}^{\pm}\dotv d\vec{l}$ are referred to as \emph{surface potentials}.  The surface potentials $f^-$ and $f^+$ define fully the fields on the inner and outer sides of $\mathcal{S}$ ($B^-$ and $B^+$) respectively

Now with \Eqnlist{Btangent}{Bcrosstangent} being implicitly satisfied, the pressure jump conditions reduce to the single condition, \Eqn{forcebalance}.

\begin{figure}
\centering
\includegraphics[width=0.4\textwidth]
 {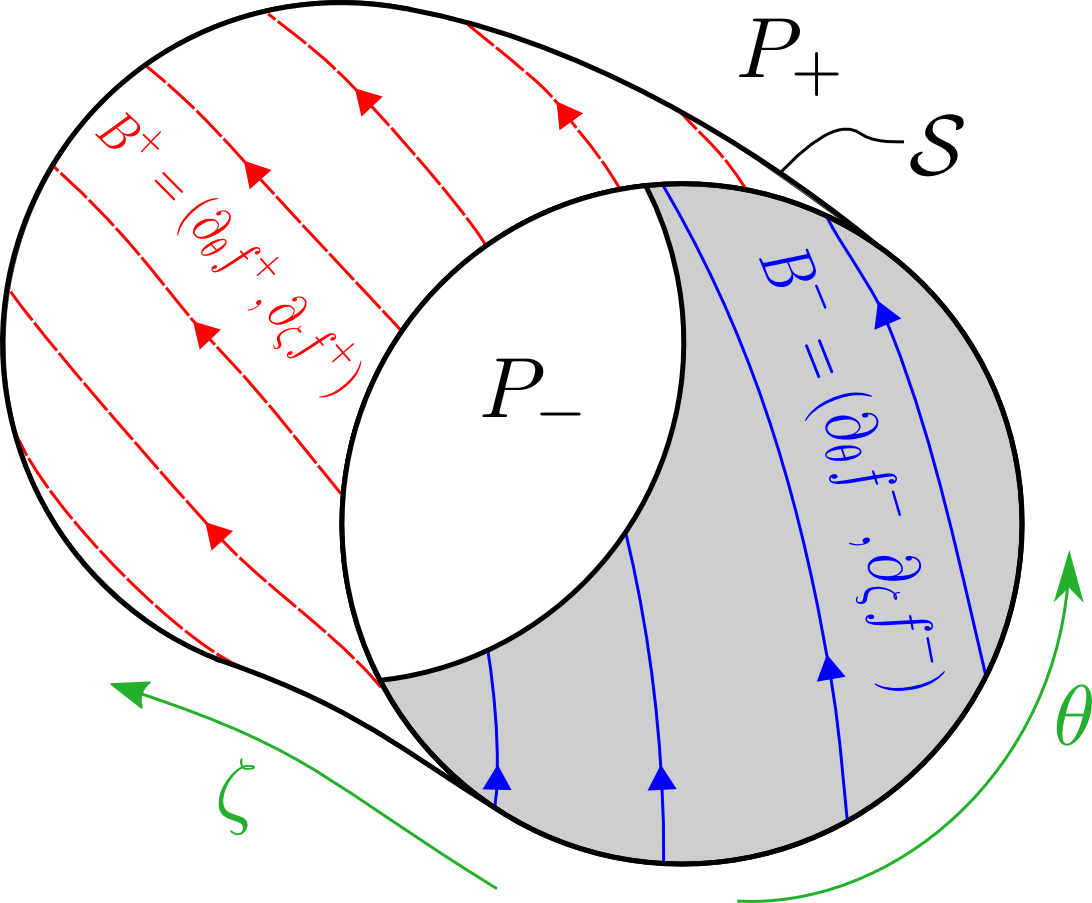}
\caption{(color online) A depiction of a toroidal segment of the surface $\mathcal{S}$ to demonstrate the properties required for the Hamiltonian.  }
\label{fig:3Dsetup} 
\end{figure}

To this end we substitute \Eqn{fdef} into \Eqn{our7} to give
\begin{align}\label{eq:our7wf}
 2 \Delta P = \sum_{i,j \in \{\theta,\zeta\}} g^{ij} \left[\partial_\theta f^{+} \partial_\zeta f^{+} - \partial_\theta f^{-} \partial_\zeta f^{-} \right]\;.
\end{align}
The goal is, given a surface (which defines $g^{ij}$) and a field on one side of the surface (which defines, say $f^-$) the goal is to find the potential on the other side of the surface ($f^+$) (See \Fig{3Dsetup}).  If $f^+$ can be found that has continuous second partial derivatives [$f^+\in C^2(P^+)$], the magnetic field will satisfy force balance and lie on the surface, satisfying \Eqnset{existence}.

The problem is symmetric, we can either, given $f^-$ find $f^+$ (work inside-out), or given $f^+$ find $f^-$ (work outside-in).  So we refer to the unknown surface field potential simply as $f$.  Then the pressure jump condition becomes a problem of calculating $f$ from the equation
\begin{align}\label{eq:HJEnot} 
 H(\theta,\zeta,\partial_\theta f, \partial_\zeta f) = \Delta P\;.
\end{align}
\Eqn{HJEnot} is a partial differential equation for $f$.  More specifically, \Eqn{HJEnot} is a time independent Hamilton--Jacobi Equation with $\partial_i f = p_i$ and Hamiltonian
\begin{align}\label{eq:7Ham} 
 H(\theta,\zeta,p_\theta,p_\zeta) = g^{ij} p_i p_j + V(\theta,\zeta)\;,
\end{align}
where there is an implicit sum of $i,j\ \in \{\theta,\zeta\}$, and $V(\theta,\zeta) = g^{ij} \partial_i f^{-} \partial_j f^{-}$ where $f^-$ has been arbitrarily chosen as the given potential.  The Hamiltonian in \Eqn{7Ham} we refer to as the \emph{pressure jump Hamiltonian}.

The solution to this autonomous Hamiltonian system can be found by solving the corresponding characteristic equations~\cite{Garabedian1964}
\begin{subequations}\label{eq:HamEq}
\begin{align}
 \dot{\theta}   &=   \pderiv[H]{p_\theta} \;, \label{eq:HamEq1} \\ 
 \dot{p_\theta} &= - \pderiv[H]{\theta  } \;, \label{eq:HamEq2} \\ 
 \dot{\zeta}    &=   \pderiv[H]{p_\zeta } \;, \label{eq:HamEq3} \\ 
 \dot{p_\zeta}  &= - \pderiv[H]{\zeta   } \;. \label{eq:HamEq4}    
\end{align}
\end{subequations}

As a Hamiltonian system, $\Delta P$ is identified with the energy and $f$ is a type two generating function (Hamilton's characteristic function) considered to generate the canonical pairs $(\theta,p_\theta)$ and $(\zeta,p_\zeta)$ through
\begin{align}\label{eq:pandf} 
 p_\theta &= \partial_{\theta} f = B_\theta\;, & p_\zeta &= \partial_\zeta f = B_\zeta \;.
\end{align}

If $f$ exists, then the Hamiltonian orbit lies on an invariant torus in phase space.  Such an orbit we refer to as \emph{regular}.

Treating the problem as Hamiltonian, one is able to utilize tools that have been developed for determination of integrability in Hamiltonian systems to investigate whether a solution ($f$) can be found that satisfies force balance.  The intent of this paper is to provide a consistent explanation to prove that inferences from the Hamiltonian system are acceptable.  

Determination of the existence of a surface potential is sufficient in the sense that it dictates definite existence criteria for the data supplied--surface shape, rotational transform, prescribed magnetic field on one side and pressure.  However, on the scale of the entire plasma system it is only a necessary condition in that many other sources of destruction may be present throughout the plasma.  These perturbations are generated outside the domain of the pressure jump Hamiltonian as it is only defined within an infinitesimal region of the surface.  

\begin{center}
\begin{table*}[t]
\hfill{}
\begin{tabular}{ll|ll}
\multicolumn{4}{c}{Physical Quantities and their Hamiltonian Equivalents} \\ \hline
 $\Delta P$ & Pressure jump & $E$ & Energy \\
 $f$ & Surface potential & $S$ &Action / Hamilton's principal function \\
 $(\theta,\zeta)$ & Curvilinear coordinates & $\vec{q}$ & Generalized coordinates \\ 
 $(B_\theta,B_\zeta)$ & Covariant components of magnetic field & $\vec{p}$ & Generalized momenta \\ 
 $(\Theta,\Phi)$ & Straight field line coordinates & $\vec{Q}$ & (Action) Angle coordinates\\ 
 $\iotabar$ & Rotational transform & $w$ & Winding number (angular frequency) \\ 
\end{tabular}
\hfill{}
\caption{A table summarizing the physical interpretations of Hamiltonian quantities in the problem.  }
\label{tab:vars} 
\end{table*} 
\end{center}

\subsection{Reduction to a $1 \frac{1}{2}$ DOF system}\label{sec:eqnsmotn} 

To simplify computation of Hamiltonian orbits, we condense these equations by dividing \Eqnset{HamEq} by \Eqn{HamEq3} to trivialize the third of Hamilton's equations and make the toroidal angle-like coordinate the ``time'' variable.  The division requires that $\dot{\zeta} \ne 0$, the implications of this being addressed in \Sec{Choiceofpm}.  

The reduced equations provide more physically relevant equations of motion.  The first describes the path of the Hamiltonian trajectory through configuration space:
\begin{align} 
\deriv[\theta]{\zeta} &= \frac{g^{\theta\theta} p_\theta + g^{\theta\zeta} p_\zeta}{g^{\theta\zeta} p_\theta + g^{\zeta\zeta} p_\zeta} = \frac{B^\theta}{B^\zeta}\;.   \label{eq:dthetadphi}
\end{align}
where \Eqnset{Bcocontra} have been utilized.  \Eqn{dthetadphi} is the equation of a field line.  Any solutions to this characteristic equation of the pressure jump Hamiltonian may correspond to field lines of the magnetic field on $\mathcal{S}$.  

The second equation of motion is
\begin{align} \label{eq:EqnMtn2} 
\deriv[p_i]{\zeta} &= \frac{\partial_{i}g^{ij} p_i p_j - \partial_{i} V}{g^{\theta\zeta} p_\theta + g^{\zeta\zeta} p_\zeta}\;, & (p_i &= B_i)\;,
\end{align}
for $i \in \{\theta,\zeta\}$.  \Eqn{EqnMtn2} shows that the canonical momentum gives the covariant components of the magnetic field along a Hamiltonian trajectory for solutions with the required rotational transform.  

The final equation of motion, \Eqn{HamEq4}, can be solved implicitly by the first two using the fact that the energy of this Hamiltonian system is conserved, i.e. the pressure jump is constant along the flux surface because neither region adjacent to our flux surface interface can support a pressure gradient.  This means $p_\zeta$ can be written as a function $p_\zeta = p_\zeta(\theta,p_\theta,\zeta;\Delta P)$.  When this is substituted into \Eqnlist{HamEq1}{HamEq2}, the entire system can be solved within two differential equations.  However, inversion of the Hamiltonian brings about an arbitrary sign, which for simplicity, we choose to be positive, discussing the choice further in \Sec{Choiceofpm}.

The system is now condensed into the differential system
\begin{subequations}\label{eq:DiffSys}
\begin{align}
 \deriv[\theta]{\zeta} & = u(\theta,\zeta,p_\theta;\Delta P) \\
 \deriv[p_\theta]{\zeta} & = v(\theta,\zeta,p_\theta;\Delta P)
\end{align}
\end{subequations}
One method of identifying the existence of $f$ is to calculate the Hamiltonian trajectories.  If a solution to this $1 \frac{1}{2}$ degree of freedom system can be found that lies on an invariant torus in $(\theta,p_\theta,\zeta)$ space with the correct winding number, the trajectory, when projected onto the 3D geometric torus, coincides with the field lines that lie on that surface.  

The solution of the pressure jump Hamiltonian is in general not unique.  It may be that many Hamiltonian trajectories are regular, and so are physical candidates for a field to satisfy force balance.  To make the solution unique one can require the corresponding field line have a certain \emph{rotational transform}, defined as
\begin{align}\label{eq:rttsfm} 
 \lim_{\Delta \zeta\rightarrow\infty} \frac{\Delta \theta}{\Delta \zeta} = \iotabar\;.
\end{align}
The pressure jump conditions, \Eqnset{existence}, allow a jump in $\iotabar$ across $\mathcal{S}$, although there is some evidence to suggest that a jump in rotational transform is unstable~\cite{Mills2009}.

Given the non-intersection of phase-space characteristics for a well-defined Hamiltonian system \cite{Arnold_1989}
and the fact that the phase-space characteristics are confined to a topological torus, it is understood that the rotational transform embodies a topological invariant of the phase-space characteristic itself. In \Sec{birk}, the existence of a map between phase-space characteristics and the magnetic field lines in configuration space that preserves characteristic topology will be shown. It is clear that since this map preserves characteristic topology, it also preserves the rotational transform.

\subsection{Action Angle Coordinates}\label{sec:actnangl} 


In action angle coordinates the canonical momenta are constant, so that the equations of motion are trivial.  The equations of motion are, after an appropriate choice of initial conditions,
\begin{align}\label{eq:AAEM} 
 \Theta &= w_\Theta t\;, & \qquad \Phi &= w_\Phi t\;.
\end{align}
where $w_\Theta$ and $w_\Phi$ are constant. Thus, 
\begin{align}
 \Theta &= \frac{w_\Theta}{w_\Phi} \Phi \;.
\end{align}
This coordinate system is, in fusion research, known as straight field line coordinates, as the magnetic field appears as a straight line in these coordinates.  Such coordinates are helpful because the rotational transform, usually a quantity that requires integration along the entire length of a field line (often infinitely long) is now explicit in the equations of motion:
\begin{align}
 \lim_{\Delta \zeta\rightarrow\infty} \frac{\Delta \theta}{\Delta \zeta} = \deriv[\Theta]{\Phi} = \frac{w_\Theta}{w_\Phi} = \iotabar\;.
\end{align}

Such coordinates can be found when $f$ (which is the generating function. cf. \Eqn{pandf}) is separable in the configuration coordinates, in this case the corresponding Hamiltonian trajectory lies on an invariant torus~\cite{Lichtenberg1992}.  

The most general form for $f$ given its definition in \Eqn{fdef} is
\begin{align}
 f = I \theta + G \zeta + \hat{f}(\theta,\zeta),
\end{align} 
where $I$ and $G$ are constants, and $\hat{f}(\theta,\zeta)$ is a function periodic in $\theta$ and $\zeta$.  The transformation to straight field line coordinates can be accomplished via the transformation
\begin{subequations}\label{eq:sflctrans}
\begin{align}
 \theta &= \Theta - \frac{\hat{f}(\theta,\zeta)}{I}, \\
 \zeta   &= \Phi\;.
\end{align} 
\end{subequations}
In the new coordinates,
\begin{align}
 f = I \Theta + G \Phi\;,
\end{align} 
and thus the magnetic field components are constant.

\section{Analytical Concerns}\label{sec:Analytical} 

\subsection{Ambiguity of Sign}\label{sec:Choiceofpm} 

The pressure jump Hamiltonian is a constraint on the square of the magnetic field, so it is expected that there are two magnetic fields that would satisfy force balance.  This arbitrariness is made explicit when one inverts the Hamiltonian to find $p_\zeta = p_\zeta(\theta,p_\theta,\zeta;\Delta P)$ in an effort to reduce the phase space.  When completing the square one has the expression
\begin{align}\label{eq:pm} 
 g^{\theta\zeta} p_\theta + g^{\zeta\zeta} p_\zeta = \pm \Delta\;,
\end{align}
where $\Delta = (g^{\zeta \zeta} \left(2\Delta P + V(\theta,\zeta) - g^{\theta\theta}p_\theta^2\right) + g^{\theta\zeta} p_\theta^2)^{1/2}$.  The left hand side of \Eqn{pm} is the result of \Eqn{HamEq3}, i.e.
\begin{align}
 \deriv[\zeta]{t} = {B}^\zeta = \pm \Delta\;,
\end{align}
showing that the two solutions correspond to magnetic fields with opposite toroidal direction.  

The information on the toroidal direction of the given field is similarly lost within the pressure jump Hamiltonian so the choice of sign must be made to reflect the initial conditions.  It is expected that physical configurations would require a field to be in the same toroidal direction on either side of an infinitely thin flux surface, otherwise there would be a very strong current sheet, susceptible to a tearing instability~\cite{Parker1990}.

The choice of sign has also been shown in Bruno and Laurence to be equivalent to choosing the sign of the rate of change of toroidal flux~\cite{B&L}.

When reducing the phase space of the Hamiltonian system, Hamilton's equations were divided by $\dot{\zeta}$, thus the condition in which $\dot{\zeta} = 0$ was necessarily lost.  A field line that does not extend toroidally corresponds to a field configuration of infinite rotational transform--a situation we ignore as it will not ergodically cover the surface, and thus will never act as a flux surface.

\subsection{Birkhoff Theorem}\label{sec:birk} 


Kaiser and Salat~\cite{K&S} solve the pressure jump discontinuity problem purely in configuration space, that is, a solution to force balance on the surface is sought that corresponds directly to geodesics covering a 3D torus.  Such an approach is limited to situations where the field within the plasma volume is zero.  However, Kaiser and Salat felt obliged to use this geodesic method because of concerns regarding the physical significance of Hamiltonian trajectories in phase space.

Kaiser and Salat's gravamen against the Hamiltonian formulation can be stated as the following:  Suppose a solution to the pressure jump Hamiltonian system is found.  This will correspond to a Hamiltonian orbit that lies in a four dimensional phase space.  The actual field line however exists on the two dimensional torus embedded in Euclidean 3-space, i.e. configuration space.  We must project a four dimensional phase space trajectory to the two dimensional configuration space--is it not possible that the projected trajectory intersects itself?

Assuming the magnetic field is nowhere zero, such intersections would make it impossible to interpret the projection as a physical field line.  

However, we will now prove that such crossings cannot occur, via a direct application of the Birkhoff theorem to \Eqnset{DiffSys}. It will thus be demonstrated that the existence intrinsic to the Hamiltonian formulation is sufficient to imply that the corresponding field line is consistent with \Eqnset{existence}.  

Our system is a $1 \frac{1}{2}$ degree of freedom Hamiltonian whose trajectories define a 2D area preserving map by integrating \Eqnset{DiffSys}.  The mappings of interest, those generated by a trajectory that lies on an invariant surface, are also twist maps as the metric is positive definite [$\det (\partial_{p_i} \partial_{p_j} H)  > 0$]~\cite{Gole2001}.

Consider the phase space variables $(q, p)$ in a 2 dimensional area preserving twist map, the Birkhoff Theorem states that, for a rotational invariant circle,~\cite{Meiss1992}
\begin{align}\label{eq:phasemap} 
 p = Y(q)\;,
\end{align}
where $Y$ is a Lipschitz function on $\mathbb{R}^2$, i.e. $Y$ satisfies
\begin{align}
\sup_{x,y\in\mathbb{R}^2}\frac{|Y(x)-Y(y)|}{|x-y|}<C\;,
\end{align}
for some bounded constant $C$.
\color{black} In this case we can write the phase space mapping generated by the Hamiltonian as
\begin{align}
 (q',p') = T(q, Y(q))\;.
\end{align}
Let us consider the operator $\pi$ that is the projection of the phase space trajectory onto configuration space,
\begin{align}
 \pi(q, p) &= q,
\end{align}
then
\begin{align}
        q' &= \pi(T[q,Y(q)]) = \alpha(q)\;.
\end{align}
Thus, as $T$ is a homeomorphism and $Y$ Lipschitz, $\alpha$ is also a homeomorphism~\cite{Meiss1992}.  The injective nature of a homeomorphism implies there will be no crossings under the mapping $\pi$.

Strictly, this is only true for a two dimensional system because the Birkhoff theorem applies only for a 2D phase space.  Some limited higher dimensional results have been found~\cite{MacKay1989}.

This means a homeomorphic mapping like \Eqn{phasemap} can be generated to define completely the evolution of the system, and the above proof applies.  Thus, when mapping the Hamiltonian trajectories to the 2D torus in configuration space no crossings are possible.  

\subsection{Existence of Invariant Tori}\label{sec:invttori} 

With the acceptance that solutions garnered from the pressure jump Hamiltonian map homeomorphically to field lines on the 3D toroidal surface, we can use the existence of the surface potential to declare that the corresponding field line satisfies \Eqnset{existence}.  

When a Hamiltonian trajectory can be transformed to action angle coordinates, the Hamiltonian orbit lies on an invariant torus; and after the mapping to configuration space the field line will lie on the surface and not intersect itself.  Conversely, if the conditions are such that the Hamiltonian trajectory is chaotic, it does not lie on an invariant torus; after the mapping the field line will not lie on the given flux surface ($\vec{B} \dotv \vec{n} = 0$ is not satisfied) and no physically consistent field can exist.  




Of great interest then, is the knowledge of whether an invariant torus of the Hamiltonian exists or not.  There are various tools to use.  Previous treatments of the problem have suggested using the KAM theory to prove the persistence of flux surfaces under small perturbation in the Hamiltonian~\cite{Berk1986}.  In the pressure jump Hamiltonian, it is not clear in advance whether a given trajectory is an invariant surface, and so the question of its persistence under perturbations is not helpful.

Practically, especially in fusion devices, the surface would be strongly perturbed and it would not be clear if the flux surface with a given rotational transform would exist.  Thus existence must be determined \emph{a posteriori}, that is, after the field line has been found, determine then whether the field line lies on an invariant surface.  The most commonly used tool to accomplish this is Green's residue criterion~\cite{Greene1979}.  Despite only partial mathematical justification,~\cite{MacKay1992} has proved helpful in investigations of this kind~\cite{Hudson2004,Paul1994,Shenker1981}.

A more computationally expensive but perhaps more physically relevant method would be the calculation of the \emph{analyticity width}~\cite{Greene1981,MacKay1992}.  Technically, it represents the domain of analyticity of the transformation to action-angle coordinates in \Eqnset{sflctrans}.  

There has been some progress in non-Hamiltonian approaches to investigate existence in problems like the pressure jump Hamiltonian.  Kaiser and Salat computationally found evidence for KAM-like behavior in their purely configurational treatment.  They developed their own purely configurational theorem to determine the extent of deformation such that no two field lines can lie on the torus and not intersect, which they termed the ``big bump'' criterion~\cite{K&S}.

\section{Conclusion}

By applying force balance to neighboring regions of plasma of finitely different pressure separated by a infinitesimally thin flux surface, three general criteria were found that must be satisfied in order for the surface to be a flux surface.  The three criteria were combined into one, referred to as the \emph{pressure jump criterion} by condition of the existence of a \emph{surface potential} $f$, a scalar function defined on the surface that is related to the magnetic field.  

A Hamiltonian--Jacobi construction of the pressure jump criterion was introduced, and the surface potential was found to play the role of the generating function important in Hamiltonian dynamics.  

The main analytical problem that faced the treatment was solved, namely the sufficiency of an existence criterion on $f$ to determine one-to-one that the corresponding field line satisfied the pressure jump criterion.  This was resolved by appealing to the Birkhoff theorem when the Hamiltonian system is integrated to form a two dimensional area preserving map.  The application of the KAM theory to Hamiltonians of this type has been questioned in other papers, so practical avenues of investigation other than the KAM theorem were suggested.

\section*{Acknowledgments}

This study was based on work done during a visit to the Princeton Plasma Physics Laboratory (PPPL), so MM and SRH would like to thank the Australian National University (ANU) and PPPL for supporting the visit.  This work was supported by the Australian Research Council and the U.S. Department of Energy Contract No DE-AC02-76CH03073 and Grant No DE-FG02-99ER54546.  The authors thank Robert MacKay for bringing the Birkhoff theorem to their attention.

{\appendix \section{Pressure Jump Condition}\label{app:presjump} 

Consider a surface $\mathcal{S}$, defined as $n=0$ where $n$ is the distance from $\mathcal{S}$.  For the purposes of deriving the pressure jump condition, the definitions of $\vec{B}^\pm$ are extended radially so they are both smooth everywhere (i.e. not by themselves discontinuous) and remain bounded in both volumes.  The discontinuous pressure profile $P(\vec{r})$ and the corresponding discontinuous complete magnetic field $\vec{B}(\vec{r})$ can then be expressed as
\begin{align}
 P       &= P_-(\vec{r}) h(-n) + P_+(\vec{r}) h(n)\;, \label{eq:pdscon} \\
 \vec{B} &= \vec{B}_-(\vec{r}) h(-n) + \vec{B}_+(\vec{r}) h(n)\;,  \label{eq:Bdscon} 
\end{align}
where $h$ is the unit Heaviside step function, which causes the discontinuity in $\vec{B}$, as both $\vec{B^\pm}$ are themselves not discontinuous.  Such a form for $P$ and $\vec{B}$ gives 
\begin{align}\label{eq:forcebalanceapp} 
 \nabla P_\pm = \vec{j}_\pm \times \vec{B}^\pm\;,
\end{align}
where $\vec{j}_\pm = \curl \vec{B}_\pm$ are the associated currents.  Dotting \Eqn{forcebalanceapp} with $\vec{B}_\pm$ gives $\vec{B}_\pm \dotv \nabla P_\pm = 0$, which implies the pressure is constant along a magnetic field line.  As a flux surface is composed of a single magnetic field line, we have the condition that on both sides of the surface
\begin{align}
 P_\pm = \textrm{const}\;.
\end{align}
Substitution of \Eqn{Bdscon} into $\div \vec{B} = 0$ gives
\begin{align}
 \div \vec{B} &= \\
         \vec{n} \dotv& \smljump{\vec{B}}\delta(n) + \div \vec{B}_- h(-n) + \div \vec{B}_+ h(n) \;,
\end{align}
where $\vec{n}$ is the unit normal to $\mathcal{S}$ and where $\smljump{x}$ is the jump of $x$ across the interface, $\smljump{x} = x_+ - x_-$.  The divergence can only be zero if
\begin{align}
 \smljump{B_n} \equiv \vec{n} \dotv \smljump{\vec{B}} = 0\;,\label{eq:Bncont} 
\end{align}
i.e., the normal component of the magnetic field must be continuous.

Similarly, when \Eqnrange{pdscon}{Bdscon} are substituted into the stress tensor in \Eqn{stressbalance}, 
\begin{align}
 \vec{n} \dotv \jump{P\vec{I} \left( \frac{1}{2} B^2 - \vec{B}\vec{B} \right)} = 0\;.\label{eq:pjvec} 
\end{align}
Substituting \Eqn{Bncont} into the above and dotting with $\vec{n}$ gives the condition
\begin{align}
 \jump{P + \frac{1}{2} B^2} = B_n \smljump{B_n} = 0\;,\label{eq:pjapp} 
\end{align}
removing the middle equality gives the \emph{pressure jump condition}.

As $B_n$ is continuous, \Eqn{pjapp} can be written as
\begin{align}
 \jump{\left(\vec{n} \cross \vec{B}\right)^2} = - 2 \smljump{P}\;,
\end{align}
so long as $\smljump{P} \ne 0$, 
\begin{align}
\smljump{\vec{n} \cross \vec{B}} \ne 0\;.\label{eq:curlcont} 
\end{align}
Crossing \Eqn{pjvec} with $\vec{n}$ gives
\begin{align}
 B_n \smljump{\vec{n} \cross \vec{B}} = 0\;.\label{eq:pjapp1} 
\end{align}
Combining \Eqn{curlcont} with \Eqn{pjapp1} implies
\begin{align}\label{eq:Btangentapp} 
 \vec{n} \dotv \vec{B}_\pm = 0\;.
\end{align}
Thus, if there is a pressure discontinuity across a surface, the field lines must lie on that surface.  

The plasma in the neighborhoods either side of the surface is assumed to be force free, so that $\nabla P = 0$.  \Eqn{forcebalanceapp} then implies that $\vec{j}_\pm$ is parallel to $\vec{B}_\pm$, which on comparison with \Eqn{Btangentapp} implies
\begin{align}\label{eq:jtangentapp} 
 \vec{n} \dotv \vec{j}_\pm = \vec{n} \dotv \left( \curl \vec{B}_\pm \right) = 0\;.
\end{align}
Thus, if there is a pressure discontinuity across a surface, the curl of the magnetic field must be parallel to the surface at all points on the surface.

 \section{Comparison of Hamiltonians}\label{app:mfhpjh} 

\begin{center}
\begin{table*}[t]
\hfill{}
\begin{tabular}{p{0.45\textwidth}|p{0.45\textwidth}}
Magnetic field line Hamiltonian & 
Pressure jump Hamiltonian \\ \hline
 $P$ implicit & $P$ explicit \\ 
 Defined throughout plasma volume & Defined on a given magnetic surface \\
 Phase space and time a representation of Euclidean 3-space & Phase space a combination of a Riemannian 2-space = magnetic surface (configuration space) and field components (momentum space) \\
All orbits are field lines & Orbits are not field lines, projections of regular orbits with specified $\iotabar$ onto configuration space are field lines
\end{tabular}
\hfill{}
\caption{A table summarizing the differences between the magnetic field line Hamiltonian and the pressure jump Hamiltonian.  }
\label{tab:mfhpjh} 
\end{table*} 
\end{center}

In this section we compare the magnetic field line Hamiltonian system often used in the literature to the pressure jump Hamiltonian system introduced in this paper.

The magnetic field in the plasma volume be written as~\cite{Boozer1983}
\begin{align}
 \vec{B}=\nabla \psi \times \nabla \theta + \nabla \zeta \times \nabla \chi\;.
\end{align}
Using the equation of the field line $d\vec{r}/dt = \vec{B}(\vec{r})$, one finds~\cite{Boozer2004}
\begin{align}
 \deriv[\theta]{\zeta} &= \phantom{-}\pderiv[\chi]{\psi} \\
 \deriv[\psi]{\zeta}   &=          - \pderiv[\chi]{\theta} \;,
\end{align}
which are of the form of Hamilton's equations.  Thus each field line can be described as the solution of the magnetic field line Hamiltonian $\chi$, in a phase space with the poloidal angle $\theta$ and toroidal flux function $\psi$ as canonical variables.  

Trajectories of this Hamiltonian correspond directly to magnetic field lines within the plasma.  Trajectories that lie on invariant tori correspond to field lines that draw out flux surfaces.  Trajectories that are chaotic correspond to chaotic field lines the compose the chaotic regions of the plasma.  

The pressure enters \emph{implicitly} into the magnetic field line Hamiltonian system, exciting currents that determine the flux functions.  The field line dynamics feeds back into the determination of the pressure as a function of position.  

In contrast, the pressure is explicit in the pressure jump Hamiltonian, but it only holds on a given toroidal surface.  While the trajectories of this Hamiltonian system are not the field lines, the projection of the phase trajectories onto the 3D torus are field lines of the given rotational transform.  If the surface potential can be found, then the surface is a flux surface.  

A summary of the differences of the Hamiltonian systems in given in \Tab{mfhpjh}.  
}


\bibliographystyle{elsarticle-num}
\bibliography{MHD}


\end{document}